\documentclass[aps,prl,reprint,groupedaddress,floatfix]{revtex4-1}
\include{SOP_header}
\begin{document}
\title{Reduced sensitivity to charge noise in semiconductor spin qubits via symmetric operation}
\author{M.~D.~Reed}
\author{B.~M.~Maune}
\author{R.~W.~Andrews}
\author{M.~G.~Borselli}
\author{K.~Eng}
\author{M.~P.~Jura}
\author{A.~A.~Kiselev}
\author{T.~D.~Ladd}
\author{S.~T.~Merkel}
\author{I. Milosavljevic}
\author{E.~J.~Pritchett}
\author{M.~T.~Rakher}
\author{R.~S.~Ross}
\author{A.~E.~Schmitz}
\author{A.~Smith}
\author{J.~A.~Wright}
\author{M.~F.~Gyure}
\author{A.~T.~Hunter}
\email{athunter@hrl.com}
\affiliation{HRL Laboratories, LLC, 3011 Malibu Canyon Road, Malibu, CA 90265, USA}
\frontmatter{We demonstrate improved operation of exchange-coupled semiconductor quantum dots by substantially reducing the sensitivity of exchange operations to charge noise. The method involves biasing a double-dot symmetrically between the charge-state anti-crossings, where the derivative of the exchange energy with respect to gate voltages is minimized.  Exchange remains highly tunable by adjusting the tunnel coupling.  We find that this method reduces the dephasing effect of charge noise by more than a factor of five in comparison to operation near a charge-state anti-crossing, increasing the number of observable exchange oscillations in our qubit by a similar factor.  Performance also improves with exchange rate, favoring fast quantum operations.}

Gated semiconductor quantum dots are a leading candidate for quantum information processing due to their high speed, density, and compatibility with mature fabrication technologies~\cite{Loss1998,Petta2005}.  Quantum dots are formed by spatially confining individual electrons using a combination of material interfaces and nanoscale metallic gates.  Although several quantized degrees of freedom are available \cite{Kim2014,Petersson2010,Culcer2012}, the electron spin is often employed as a qubit due to its long coherence time \cite{Wang2013,Veldhorst2014b}.   Spin-spin coupling may be controlled via the kinetic exchange interaction, which has the benefit of short range and electrical controllability.  Numerous qubit proposals use exchange, including as a two-qubit gate between ESR-addressed spins~\cite{Veldhorst2015}, a single axis of control in a two dot system also employing gradient magnetic fields \cite{Brunner2011} or spin-orbit couplings \cite{Nowack2007}, or as a means of full qubit control on a restricted subspace of at least three coupled spins~\cite{DiVincenzo2000,Medford2013,Eng2015}.  However, since exchange relies on electron motion, it is susceptible to electric field fluctuations, or charge noise.  Limiting the consequence of this noise is critical to attaining performance of exchange-based qubits adequate for quantum information processing.

\begin{figure}[h!]
\includegraphics{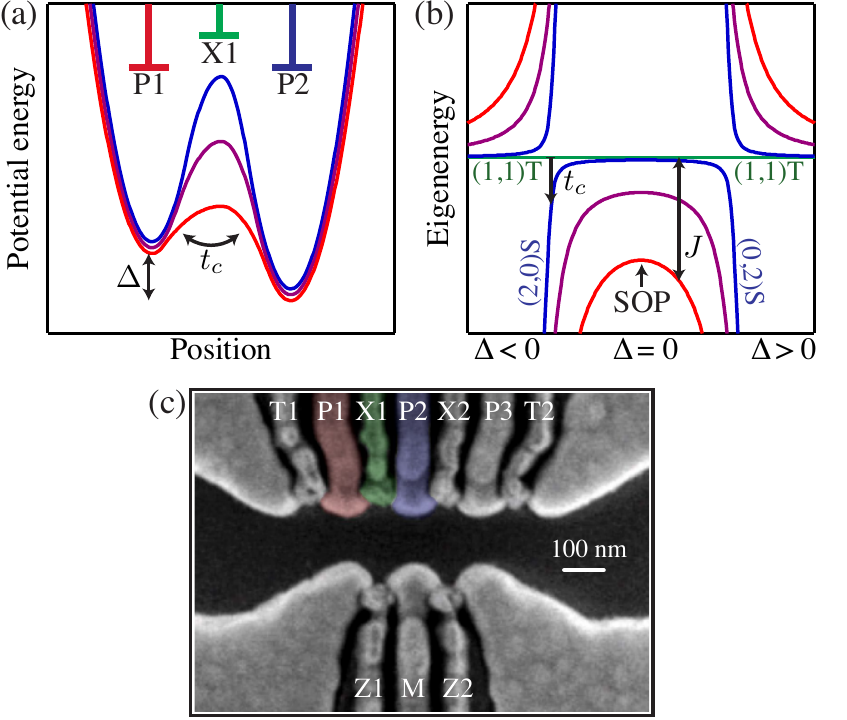}
\caption{(a) Schematic of a double-dot potential energy as a function of position, in which the left dot has a chemical potential higher than the right dot by the detuning $\Delta$.  The chemical potentials are predominantly controlled by gates P1 and P2 (schematized above each) and the barrier by X1.  When X1 is forward-biased, the tunnel coupling $\ts{t}{c}$ is increased, with the blue curve indicating low $\ts{t}{c}$ and the red curve high $\ts{t}{c}$. (b) Schematic of eigenenergies of the double-dot as a function of $\Delta$, according to a simple Hubbard model.  Here the singlet states are again colored blue to red as $\ts{t}{c}$ is increased, while the triplet states are colored green.  Exchange $J$ is the difference between singlet and triplet energies; $\partial J/\partial\Delta$ vanishes at the symmetric operating point $\Delta=0$.  Note that our definition of $\Delta$ corresponds to the chemical potential difference of singly occupied dots and not distance from the anti-crossing (typically notated $\varepsilon$ as in Refs.~\onlinecite{Petta2005, Maune2012, Eng2015}). (c) Representative electron micrograph of a triple quantum dot, with gate traces labeled.  We false-color gates P1, X1, and P2 respectively red, green, and blue.}
\label{cartoon}
\end{figure}

Charge noise in semiconductor quantum dots may originate from a variety of sources including electric defects at interfaces and in dielectrics~\cite{Paladino2014}.  These defects typically result in electric fields that exhibit an approximate $1/f$ noise spectral density.  Conventional routes for reducing charge noise include improving materials and interfaces~\cite{Borselli2015} and dynamical decoupling \cite{Wang2014, Cerfontaine2014, BrownBiercuk2014, DasSarma2014}.  In this Letter, rather than addressing the microscopic origins or detailed spectrum of charge noise, we introduce a ``symmetric'' mode of operation where the exchange interaction is less susceptible to that noise.  This is done by biasing the device to a regime where the strength of the exchange interaction is first-order insensitive to dot chemical potential fluctuations but is still controllable by modulating the inter-dot tunnel barrier.  This dramatically reduces the effects of charge noise.

The principle of symmetric operation can be understood by treating charge noise as equivalent to voltage fluctuations on confinement gates.  This approximation is valid when interfaces proximal to gates are the dominant source of noise \cite{Paladino2014}.  In this context, noise sensitivity may be reduced by biasing the device to a ``sweet spot'' where small changes in gate voltages only weakly alter the strength of the exchange interaction.  Previously explored methods include using a triple quantum dot with balanced exchange interactions~\cite{Taylor2013, Fei2015} (see the Supplementary Material for a comparison), operating far from the (1,1) charge regime where excited states flatten the exchange profile \cite{Dial2013, Bluhm2015, Fei2015}, using double-dots populated with more than two electrons~\cite{Higginbotham2014}, or tailoring exchange derivatives via magnetic field gradients~\cite{Wong2015}.  The strategy we pursue in this Letter has the advantage of employing only pairwise exchange without requiring high or inhomogeneous magnetic fields and maintains tunability of the exchange coupling rate from being negligibly small to many GHz.

Symmetric operation is diagrammed in \reffig{cartoon}.  The difference in chemical potential between two dots is denoted $\Delta$ and is predominantly controlled by two gates labeled P1 and P2 in \reffig{cartoon}(a).  For an ideal double quantum dot, $\Delta=\alpha(\ts{V}{P1}-\ts{V}{P2})$ where $\alpha$ is the ``lever arm'' that converts voltage to chemical potential.  A third gate, labeled X1, controls the potential barrier that sets the rate at which an electron can tunnel, $\tc/h$.  Figure 1(b) shows eigenenergies for a double dot as calculated with a Hubbard model.  Crucially, although the detuning $\Delta$ is often used to control $J$, the tunnel coupling $\tc$ can also modify the energy difference between the singlet and triplet energy eigenstates, $J(\Delta, \tc)$.  In particular, $J(\Delta=0, \tc)$ is a ``sweet spot'' where the effects of charge noise on exchange are reduced because $\partial J/\partial \Delta=0$ \cite{Taylor2007,Klinovaja2012}, as evident from \reffig{cartoon}(b).  We refer to $J(\Delta=0, \tc)$ as a symmetric operating point (SOP) because the double quantum dot is biased to the center of the (1,1) charge cell and equidistant from the (2,0) and (0,2) anti-crossings.  

Although any system of exchange-coupled quantum dot pairs could potentially benefit from symmetric operation, we use Si-based triple-quantum-dot devices for our demonstration.  A SEM image of a device is shown in \reffig{cartoon}(c).  Metallic plunger gates P1-P3 are deposited on an undoped Si/SiGe heterostructure.  When the plungers are forward biased, individual electrons are drawn from a bath and accumulate beneath the plungers \cite{Eng2015, Borselli2015}.  The X and T gates are deposited on an insulating layer that overlaps the plungers and control tunnel barriers between the dots and to the electron bath.  Some devices in our study differ from Ref.~\onlinecite{Eng2015} by the addition of a metal screening gate which prevents charge accumulation under gate leads \cite{Zajac2015}.  A proximal dot charge sensor formed by the M and Z gates enables single-shot readout of the qubit state \cite{Eng2015}.  P and X gate control lines are capable of nanosecond pulse rise times and amplitudes of $140\mV$.  The devices are operated in a dilution refrigerator, giving $T_{e}\sim100\mK$.

The third dots in our devices enable initialization and measurement (see Fig. 2(a) of Ref.~\onlinecite{Eng2015}).  In the experiments described below, we prepare the qubit state by biasing near the (1,0,1)-(1,0,2) charge transition where a two-electron singlet state is preferentially loaded into dot 3.  One of the electrons is then transferred into dot 2 by ramping P2 and P3.  We define this singlet state between dots 2 and 3 as the north pole of a qubit Bloch sphere \cite{DiVincenzo2000}.  Exchange between dots 1 and 2 occurs at a frequency $J(\Delta,\tc)/h=J(\vec{V})/h$, where $\vec{V}$ denotes the gate voltages.  This interaction rotates the qubit state about an axis which is tipped 120$^\circ$ from the north pole \cite{DiVincenzo2000, Medford2013, Eng2015}.  After some evolution, we measure the qubit state using Pauli blockade by biasing near the (1,0,2)-(1,1,1) charge transition.  Sweeping the evolution time yields Rabi oscillations which have a maximum contrast of 75\% due to the tilted rotation axis.

\begin{figure}
\includegraphics{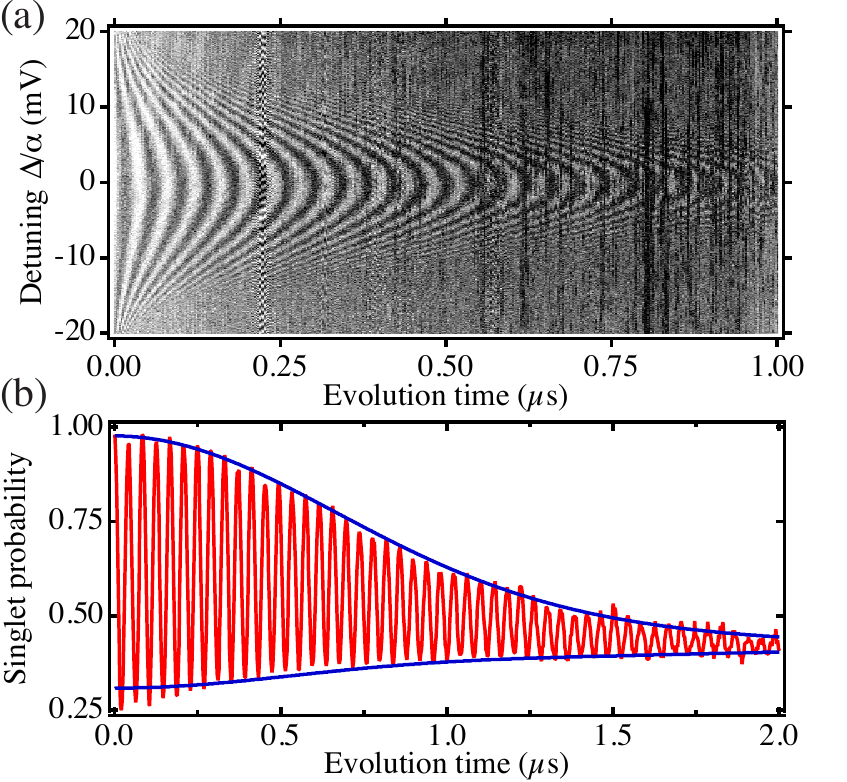}
\caption{Rabi oscillations at a SOP. (a) We observe exchange oscillations by biasing to a detuning ($y$-axis) for a given time ($x$-axis).  The middle of the plot, where $\Delta=0$ and $J$ is minimized, corresponds to the SOP.  The grayscale measures singlet probability and ranges from 100\% (white) to 25\% (black).  (b) Evolving at $\Delta=0$ reveals a time-domain Rabi oscillation showing a double Gaussian decay with a 1/e time of $1.0~\mu\mathrm{s}$ due to hyperfine interactions and $1.5~\mu\mathrm{s}$ due to charge noise.}
\label{rabi}
\end{figure}

We demonstrate singlet-triplet Rabi oscillations in \reffig{rabi}(a) by sweeping the exchange duration and $\Delta$ while holding $\tc$ constant.  The Rabi frequency is given by $J(\Delta,\tc)/h$ and is large even with $\Delta=0$ because $V_{\mathrm{X1}}$ is forward-biased during evolution, increasing $\tc$.  $J$ increases with $|\Delta|$, producing a chevron pattern.  The number of resolvable oscillations is greatest at the SOP ($\Delta=0$), giving preliminary indication that using a SOP can enhance the quality of the exchange interaction.  This improvement can be interpreted in the context of gate-referred charge noise.  As discussed in the Supplementary Material, for large detuning $|dJ/d\Delta|\approx J^2/\tc^2$.  Thus, as $J$ is increased by detuning, it becomes quadratically more susceptible to charge noise.  When $\Delta=0$, however, the dominant derivative is now $dJ/d\ts{V}{X1}=(\partial J/\partial \tc)(d\tc/d\ts{V}{X1})$, proportional only to $J$.  Increasing $J$ with $\tc$ then only linearly increases susceptibility to charge noise.   (This scaling is valid when $J$ is exponentially dependent on $\ts{V}{X1}$; we later show that it can be {\it sub}-exponential and thus even more favorable.) 

The shape of the Rabi oscillations shown in \reffig{rabi}(b) can be modeled with a two-channel decay process.  One process is due to the hyperfine interaction between the electron spin and that device's natural abundance of $^{29}$Si nuclei and is described by Eq. 12 of Ref.~\onlinecite{Ladd2012}.  The other process is due to charge noise, which, for the $1/f$ spectrum of noise seen in these devices \cite{Eng2015}, imposes a Gaussian envelope.  For this illustrative example, the relatively low value of $J$ and the natural isotopic abundance of this sample renders the charge decoherence time comparable to the magnetic dephasing time.  In the discussion that follows, however, because we focus on higher values of $J$ in isotopically purified silicon samples, charge noise will dominate the decay envelope.

For gate-referred $1/f$ charge noise, this envelope can be expressed as $\exp(-\sigma_{\mathrm{V}}^2\sum_j |dJ/dV_j|^2 t^2/\hbar^2)$.  Here, $\sigma_{\mathrm{V}}^2$ is the variance of the noise (e.g. the noise spectral density integrated over relevant timescales) and $j$ indexes all gates; see Ref.~\onlinecite{Dial2013} and the Supplementary Material.  Increasing the Rabi decay time for fixed $J$ can then be accomplished by reducing $\sum_j |dJ/dV_j|^2$ \cite{Taylor2007}.  We define the insensitivity $\mathcal{I}$ as
\begin{equation}
\mathcal{I} = {J}/{\sqrt{{\sum}_j |dJ/dV_j|^2}},
\label{insensitivitydef}
\end{equation}
which has units of voltage.  With this metric, the expected envelope of Rabi oscillations is $\mathrm{exp}\left[-(Jt/\hbar)^2(\sigma_{\mathrm{V}}/\mathcal{I})^2\right]$, so that the number of oscillations that occur before the amplitude decays by $1/e$ is $\mathcal{I}/(2\pi\sigma_{\mathrm{V}})$.  As long as $\sigma_{\mathrm{V}}$ is not too strong a function of control parameters (e.g. noise not varying from one gate to the next), we can optimize device performance by maximizing $\mathcal{I}$ with respect to $\vec{V}$.  In particular, only the charge noise variance and not the detailed structure of its spectral density is relevant to this calculation, enabling predictions of bias-dependent charge noise performance based on device electrostatics.

\begin{figure}
\includegraphics{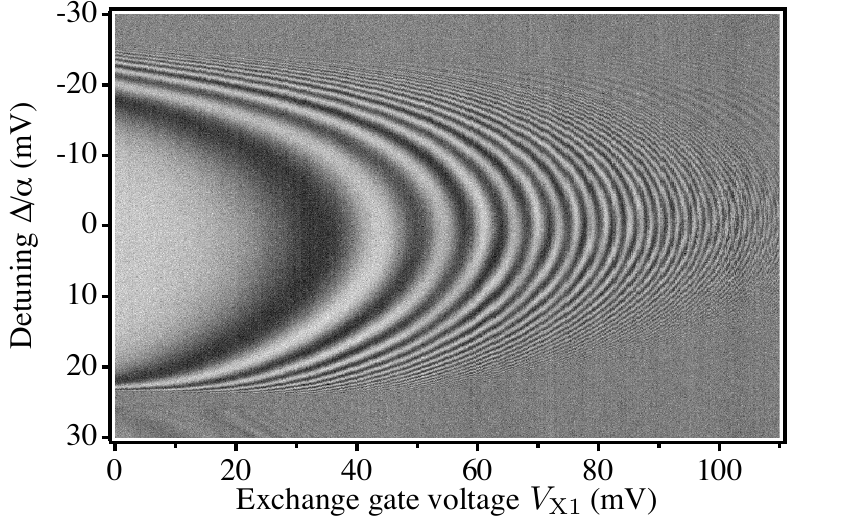}
\caption{``Fingerprint'' plot demonstrating the dependence of exchange on $\Delta$ and $V_{\mathrm{X1}}$.  We plot the average singlet probability after evolving for $500\ns$ at a potential specified by the axes.  The $z$-scale is the same as \reffig{rabi}.  A faint set of additional fringes is present in this data (prominent near $(\Delta,V_{\mathrm{X1}})=(-20,100))$, likely due to excited state population (Supplementary Material).  The device used here and in all subsequent figures differs from the device used in \reffig{rabi} by the addition of a screening gate \cite{Zajac2015} and the use of enriched $^{28}\mathrm{Si}$ ($800~\mathrm{ppm} ~ ^{29}\mathrm{Si}$)\cite{Eng2015}.  }
\label{fingerprint}
\end{figure}

In order to demonstrate the advantage of symmetric operation, we must independently control $\Delta$ and $\tc$.  The plunger and exchange gates affect both parameters due to capacitive cross-talk.  Using a routine described in the Supplementary Material, we orthogonalize these control axes.  Modulation of $\tc$ is accomplished by changing $V_{\mathrm{X1}}$ along with small compensating voltages on plunger gates, while $\Delta$ is modified by biasing P1 and P2 with approximately equal and opposite voltages.  We show the effect of these parameters on $J$ in \reffig{fingerprint}, where we evolve for a fixed time at a point that is swept in both $\tc$ (parametrized by $V_{\mathrm{X1}}$) and $\Delta$.  This ``fingerprint'' plot shows fringes due to varying $J$, the curvature of which indicates where $dJ/d\Delta=0$.  This locus of points, which on this plot is approximately parallel to the $x$-axis due to our orthogonalization scheme, is known as the symmetric axis and indicates the location of the SOP for a given $J$.

\begin{figure}
\includegraphics{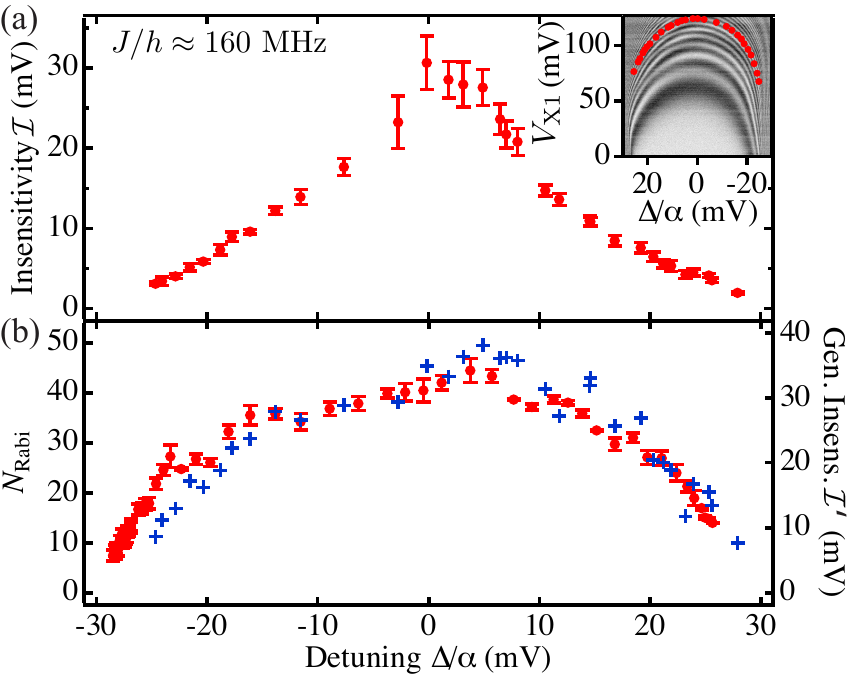}
\caption{Insensitivity and number of fringes along a constant-$J$ curve. (a) We measure $\mathcal{I}$ along a contour where $J/h\approx160\mhz$ for various $\Delta$ and $V_{\mathrm{X1}}$ (inset).  $\mathcal{I}$ is peaked near $\Delta=0$ at a value of $\mathord{\sim}30\mV$, rapidly falling to below $5\mV$ as $|\Delta|$ is increased.  (b) We verify that $\mathcal{I}$ is a valid proxy for device performance by measuring the number of fringes present in a $1/e$ decay time with a series of time-Rabi experiments where the evolution point is swept along the same contour.  Due to the presence of two evolution frequencies in this device (Supplementary Material), we first apply a high-pass filter to the time-domain data before extracting the decay coefficient.  We plot the product of that coefficient and $J/h$ as closed red circles.  We also plot a generalized definition of $\mathcal{I}$ (blue crosses) which better models the data (Supplementary Material).}
\label{insensitivity}
\end{figure}

Symmetric operation maximizes $\mathcal{I}$.  To demonstrate this, we choose various combinations of $\Delta$ and $\tc$ where $J/h=160\mhz$,  shown in the inset of \reffig{insensitivity}(a).  At each evolution point, we explicitly measure $\mathcal{I}$ by determining how the Rabi oscillation frequency changes due to small perturbations in each of the seven relevant gate voltages.  We plot the resulting insensitivity in \reffig{insensitivity}(a) and find that it is maximized at $\Delta=0$ with a value of $\mathord{\sim}30\mV$ and rapidly decreases to less than $5\mV$ for large $\Delta$.

To test the validity of $\mathcal{I}$ as a metric for the effects of charge noise, we measure the number of Rabi oscillations $N_{\mathrm{Rabi}}\equiv J\tau/h$ that occur in a 1/e decay time $\tau$.  If the gate-referred charge-noise variance $\sigma_{\mathrm{V}}^2$ were equal and uncorrelated on all gates, then we would expect $\mathcal{I}\propto N_{\mathrm{Rabi}}$.  To determine $N_{\mathrm{Rabi}}$, we measure $\tau$ along the voltage arc where $J(\Delta, \tc)/h=160\mhz$.  The resulting $N_{\mathrm{Rabi}}$ is plotted in \reffig{insensitivity}(b).  Though it qualitatively follows $\mathcal{I}$ and is maximum near $\Delta=0$, the quantities are not strictly proportional, indicating that our assumptions are not fully supported.  In particular, as discussed in the Supplementary Information, by including some knowledge of the relative geometries of the P and X gates in this device, we can more accurately model $N_{\mathrm{Rabi}}$ with a generalized definition of $\mathcal{I}$ (blue crosses in \reffig{insensitivity}(b)).

\begin{figure}
\includegraphics{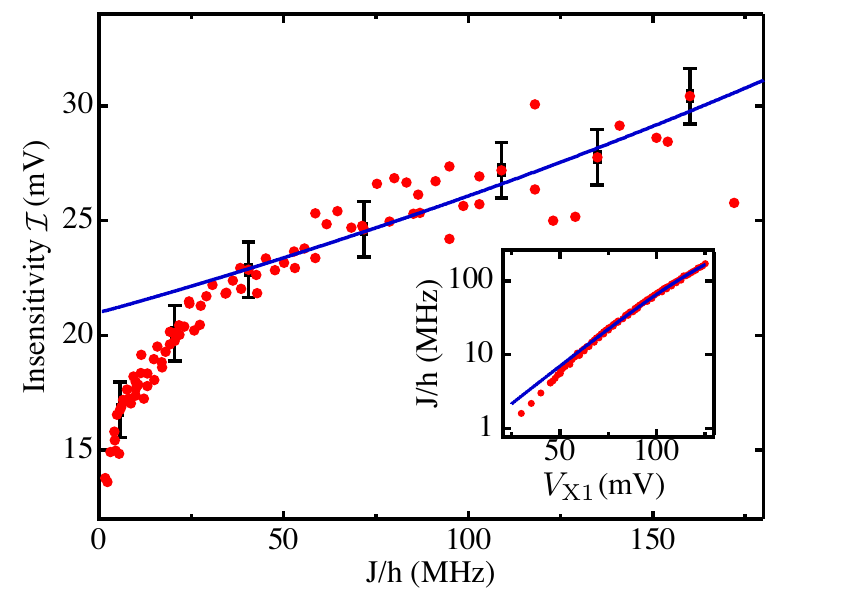}
\caption{Asymptotic behavior along the symmetric axis.  (inset) We measure $J$ as a function of $V_{\mathrm{X1}}$ along the symmetric operating vector, $J(\Delta=0,\tc)$, and find that it is sub-exponential at large $J$.  (main figure) As a consequence of this asymptotic behavior, $\mathcal{I}$ monotonically increases with $J$, roughly doubling over the relevant voltage range.  We plot several representative error bars in black and suppress the rest for clarity.  We compare the insensitivity of several similar devices in the Supplementary Material.}
\label{IvsJ}
\end{figure}

Charge noise is not the only source of degradation for exchange-based control: finite $T_1$ and hyperfine dephasing will also play a role.  Their effects are generally diminished by making the evolution time as short as possible, requiring $J$ to be maximized.  When $J$ is controlled solely by $\Delta$, this poses a major problem as $\mathcal{I}$ will drop precipitously; one must trade-off between infidelity sources.  Fortunately, for symmetric operation there is no such penalty; indeed, performance actually improves.  We see this by first noting that $J$ will depend exponentially on the height of the tunnel barrier when $\tc$ is small.  As we approach the large-$J$ limit, however, the SOP double-dot merges into a larger two-electron single dot where the former barrier is shallow.  In this regime, $J$ will saturate to near that broader potential's orbital excited energy and not depend on $\tc$ (Supplementary Material).  This is reflected in the inset of \reffig{IvsJ}, where $J(\vec{V})$ behaves sub-exponentially with increasing $V_{\mathrm{X1}}$.  Since the main contributor to $\mathcal{I}$ when $\Delta=0$ is this exchange gate, we see in \reffig{IvsJ} that $\mathcal{I}$ monotonically increases with $J$, doubling over the selected range.  We may fit this behavior at high-$J$ using a one-dimensional WKB approximation appropriate for shallow barrier tunneling \cite{RauModifiedWKB} which correctly captures the sub-exponential behavior of $J$ and increased $\mathcal{I}$ but breaks down at low $\tc$.  In some cases, detailed 3D single-electron Poisson-Schr\"odinger simulations including disorder captures the full behavior.

We have shown that symmetric operation improves the noise characteristics of semiconductor qubits employing the exchange interaction.  Compared to detuning, it is substantially less sensitive to noise, particularly for large $J$ where fast gates are possible.  Though we demonstrated symmetric operation with an exchange-only Si triple dot, the principle should work equally well in any device where $\Delta$ and $\tc$ can be separately controlled.  Indeed, we became aware of recent demonstrations in GaAs double dots during the preparation of this manuscript \cite{Bertrand2015, Martins2015}.  Future work will be to characterize the benefits of symmetric operation on control fidelity using techniques such as randomized benchmarking.  

We thank C. Jones, J. Kerckhoff, S. Meenehan, and D. Underwood for discussions.  This research was developed with funding from the Defense Advanced Research Projects Agency (DARPA). The views, opinions, and/or findings contained in this material are those of the authors and should not be interpreted as representing the official views or policies of the Department of Defense or the U.S. Government.

\end{document}


\beginsupplement
\widetext

\title{Supplemental Materials: Reduced sensitivity to charge noise in semiconductor spin qubits via symmetric operation}
\author{M.~D.~Reed}
\author{B.~M.~Maune}
\author{R.~W.~Andrews}
\author{M.~G.~Borselli}
\author{K.~Eng}
\author{M.~P.~Jura}
\author{A.~A.~Kiselev}
\author{T.~D.~Ladd}
\author{S.~T.~Merkel}
\author{I.~Milosavljevic}
\author{E.~J.~Pritchett}
\author{M.~T.~Rakher}
\author{R.~S.~Ross}
\author{A.~E.~Schmitz}
\author{A.~Smith}
\author{J.~A.~Wright}
\author{M.~F.~Gyure}
\author{A.~T.~Hunter}
\email{athunter@hrl.com}
\affiliation{HRL Laboratories, LLC, 3011 Malibu Canyon Road, Malibu, CA 90265, USA}
\frontmatter{}

\section{Formal Definitions for Charge Noise and Insensitivity}
In this section we formalize how we quantify charge noise and insensitivity.  Charge noise may originate from a variety of sources in a sample.  These sources each create electrostatic potentials which sum at each location $\vec{r}$, resulting in a fluctuating electric potential $\delta\phi(\vec{r},t)$, which we assume is \emph{not} a function of applied voltages.  This  potential is added to the potential generated by our gates, $\phi[\vec{r};\vec{V}(t)]$ which is controlled by the voltages applied to each gate.  These voltages are notated as a time-dependent vector in voltage space, $\vec{V}(t)$, with a dimension given by the number of gates, $\ts{N}{gates}$.  We model the effect of $\delta\phi(\vec{r},t)$ by imagining a fluctuating voltage vector, $\vec{v}(t)$, such that
\be
\phi[\vec{r};\vec{V}+\vec{v}(t)] = \phi(\vec{r};\vec{V})+\vec{v}(t)\cdot\vec{g}(\vec{r}) \approx \phi(\vec{r};\vec{V})+\delta\phi(\vec{r},t),
\ee
where $\vec{g}(\vec{r})$ is the voltage-gradient of the potential with components $g_{j}(\vec{r})=d\phi(\vec{r};\vec{V})/dV_j$.  A vector $\vec{v}(t)$ which closely satisfies this approximation comes from the spatially-averaged Moore-Penrose pseudoinverse
\be
\label{vdef}
\vec{v}(t) = \biggl[\int \vec{g}(\vec{r}')\vec{g}^\dag(\vec{r}')d^3\vec{r}'\biggr]^{-1}\cdot\int  \delta\phi(\vec{r},t)\vec{g}^\dag(\vec{r})d^3\vec{r},
\ee
where the spatial integral for averaging should be well-chosen around the active quantum dot region.  This notional pseudoinverse accomplishes the projection of charge-noise induced fields onto gate-induced fields, and formalizes what is referred to in the main text as ``gate-referred charge noise."  Obviously, $\vec{v}(t)$ will not capture all effects of charge noise, since in particular it is possible for charge-noise-induced electric field fluctuations to occur in a direction orthogonal to the electric fields produced by the gates.  However, we do not have experimental access to $\delta\phi(\vec{r},t)$, so we instead focus our attention on the effect of fluctuating gate voltages.

Insensitivity compares the exchange energy $J$ to the variation of $J$ with respect to changes in gate voltage.  We desire $J$, as observed in an experiment observing ensemble-averaged coherent oscillations, to be large and the variation of $J$ to be small. Insensitivity may be regarded as a figure of merit indicating how many exchange fringes are expected to be observed in a decaying oscillation per unit of root-mean-square voltage noise.  In general, fringe decay may be modeled via the envelope function
\be
G(t) = \exp\biggl[-\int_0^\infty \frac{d\omega}{\pi(\hbar\omega)^2} F(\omega t) S_J(\omega)\biggr],
\label{Gdef}
\ee
where $S_J(\omega)$ is the noise spectral density for fluctuations in $J$ and $F(\omega t)$ is the unitless filter function for the experiment which depends on pulse shapes and sequences.

For exchange-coupled quantum dots, $J$ is a function of applied voltages $\vec{V}$, and we add to these voltages the fictitious fluctuations $\vec{v}(t)$ discussed above in \refeq{vdef}.  Charge noise is therefore expected to modulate $J(\vec{V})$ as
\be
J(\vec{V}+\vec{v}(t))\approx J(\vec{V})+\sum_{j=1}^{\ts{N}{gates}}v_j(t)
    \frac{dJ(\vec{V})}{dV_j}.
\ee
To arrive at $S_J(\omega)$, we consider the ensemble-averaged correlation function
\be
\langle J(\vec{V}+\boldsymbol\delta(t)) J(\vec{V})\rangle - J^2(\vec{V}) = \sum_{jk} \frac{dJ}{dV_j} \langle v_j(t)v_k(0)\rangle \frac{dJ}{dV_k}.
\ee
We are assuming that voltage fluctuations $v_k(t)$ are zero-mean and stationary.  In general, charge noise will appear to come from one gate more than another, so the autocorrelation functions $\langle v_j(t)v_j(0)\rangle$ will vary with gate index $j$.  Moreover, the gates do not have orthogonal action, and so gate-to-gate correlations are to be expected.  However, time-correlations are expected to be roughly independent of the gate.  We separate out gate-to-gate correlations and time-correlations with a couple of definitions.  Define
\be
C_{jk} = \ts{N}{gates}\frac{\langle v_j(0) v_k(0)\rangle}{\sum_\ell \langle v_\ell(0) v_\ell(0) \rangle},
\ee
a unitless matrix whose trace is the number of gates, $\ts{N}{gates}$.  We may then define our formal average noise spectral density as
\be
S_V(\omega) = \frac{1}{\ts{N}{gates}}\sum_{jk} [C^{-1}]_{kj} \int_{-\infty}^\infty  \langle v_j(t) v_k(0)\rangle \cos(\omega t) dt.
\ee
In an actual Rabi experiment we witness an effective average voltage variance of
\be
(\sigma_V t)^2 \approx \frac{2}{\pi}\int_{1/T}^\infty \frac{d\omega}{\omega^2}S_V(\omega) F(\omega t).
\ee
Here the lower limit of the frequency integral is set to $1/T$, where $T$ is roughly the averaging time of the experiment.  This lower limit is required for convergence of the integral when $S_V(\omega)$ has a $1/f$ character.  The filter function for the Rabi pulse ($F(z)=\sin^2(z/2)$, if the pulse is square) provides the high-frequency cut-off (and an extra logarithmic dependence to the pulse-time dependence which we neglect).  Applying \refeq{Gdef}, we therefore expect decay of the form
\be
\begin{split}
G(t) &= \exp\biggl[-\frac{\sigma_V^2 t^2}{2\hbar^2}\sum_{jk} \frac{dJ}{dV_j}C_{jk}\frac{dJ}{dV_k}\biggr]
\\
     &= \exp\biggl[-\left(\frac{Jt}{\hbar}\right)^2\left(\frac{\sigma_V}{\mathcal{I}}\right)^2\biggr].
\end{split}
\ee
An appropriate definition of the insensitivity would then be
\be
\mathcal{I} = \frac{J}{\sqrt{\sum_{jk} (\partial J/\partial V_j)C_{jk}(\partial J/\partial V_k)}}.
\ee
Unfortunately, the correlation matrix $C_{jk}$ is not directly available from experimental measurements.  We therefore typically use an effective insensitivity which presumes that $C_{jk}$ is the identity matrix of dimension $\ts{N}{gates}$.  This definition of $\mathcal{I}$ depends on defining the voltage noise variance $\sigma_V^2$ under similar assumptions as above, emphasizing that care must be taken in comparing insensitivities between devices with differing gate geometries.

\section{Geometrically Informed Insensitivity}

We may improve upon the definition of insensitivity given in the main text by relaxing the assumption that $C_{jk}$ in Eq. (S10) is the identity matrix.  In particular, this may account for the fact that the exchange gates in our devices are both physically smaller and more distant from the 2DEG than the plunger gates.  This is an issue because we use the derivative of $J$ with respect to each gate as a proxy for the effect of noise on that gate's control axis and these derivatives will be artificially reduced for exchange gates by the aforementioned effects.  Since we believe our charge noise originates in the oxide barrier below the gate stack (and not actually noise on the physical gates), our assumption that we can model our charge noise as a fixed noise variance on all gates is violated.  

In order to more accurately model exchange noise, we scale the measured derivative of $J$ with respect to those gates to account for these physical differences.  Equivalently, we may instead divide the plunger gate derivatives by the same factor.  Effectively, this is a choice of what gate layer we refer our charge noise to.  This choice maintains approximate agreement between the conventional value of $\mathcal{I}$ and this generalized version at the SOP, since the derivatives of the plunger gates play only a small role in $\mathcal{I}$ at that bias point.  As shown in Fig. 4(b) of the main text, we can get very good agreement between our measured number of exchange oscillations (red circles) and a generalized insensitivity definition (blue crosses) by scaling the plunger gate derivatives by a factor of $1/4$ relative to the exchange gates.  Formally, following Eq. (S10), we define 
\begin{equation}
	C_{jk} = A_j^2 \delta_{jk},
\end{equation}
and set $A_{\rm P} = 1/4$ (e.g. for plunger gates) and $A_{\rm X} = 1$ (exchange gates) to best model our data.

We may qualify this factor of four with electronic structure simulations of our devices. These simulations can be performed at various levels of approximation, from classical electrostatics to fully quantum mechanical. The classical electrostatics calculation considers the response to sources of charge noise on the confining electrostatic potential $\varphi(\vec x)$ that supports the few-electron quantum dot states. In these simulations, it is convenient to define a set of nominally orthogonal ``natural'' coordinates that characterize the principle deformations of a double well potential in which a two-electron system resides.  We define detuning $\varepsilon \equiv \varphi_{\rm left} - \varphi_{\rm right}$ and barrier height $\chi \equiv \varphi_{\rm barrier} - \left( \varphi_{\rm left} + \varphi_{\rm right} \right) / 2$, where $\varphi_{\rm left, right, barrier}$ refer to the potential values in the centers of the left and right dot and in the barrier separating them.  In general, charge noise will influence these deformations differently.  We define the noise energy as the root-mean-square fluctuation of the respective potential deformations $\upsilon_{\nu} \equiv \sqrt{ \left< \delta \nu^2 \right> }$ with $\nu \in \{\varepsilon,\chi\}$.

The response of the electrostatic potential to charge noise, associated with dipolar fluctuators in the gate dielectrics and their associated material interface, is computed via an inverse-Green's function technique.  This takes the entire metal-dielectric-semiconductor device structure into account and allows simultaneous evaluation of the contributions from each material. For the device studied in this paper, we find that barrier height noise is smaller than detuning noise, with $\upsilon_\varepsilon \simeq 2.5 \, \upsilon_\chi$.  This validates the experimentally observed benefit of the symmetric operating point.  We may relate this factor to measurable quantities (e.g. gate-referred noise voltages and insensitivities) using the calculated lever arms that relate electrostatic potential changes to individual gate voltages.  As discussed above, for our devices the relatively smaller size of the physical exchange gate translates into a much weaker lever arm (by a factor of $\sim\!7$) compared with the plunger gates.  Our simulations therefore yield for the ratio of gate-voltage-referred P and X gate noise, $\sigma_{\rm V_P} / \sigma_{\rm V_X} \simeq 0.25$.  This is in excellent agreement with the factor chosen for generalized insensitivity to best model our measurement of $N_{\rm Rabi}$ vs. $\Delta/\alpha$ in Fig. 4(b) of the main text.

\section{Model for Exchange and Insensitivity vs. Voltage}
In this section we give a simple analytic model for the exchange energy as a function of control gate biases,  $J({\bf V})$, derived from a modified WKB approximation and the tight-binding model.   It is well-known that $J$ can be modeled as the difference of the lowest energy eigenstates of the two-electron Hubbard Hamiltonian
\begin{multline}
	H = \sum_{j=1,2; \sigma}{\mu_j c^\dagger_{j,\sigma} c_{j,\sigma}^\nodag + U n_{j,\sigma}\left(n_{j,\sigma}-1\right)} \\
+
	\frac{t_c}{\sqrt{2}} \left(c_{1,\sigma}^\dagger c_{2,\sigma}^\nodag + c_{2,\sigma}^\dagger c_{1,\sigma}^\nodag \right),
\label{hamiltonian}
\end{multline}
parameterized by the tunnel coupling $\ts{t}{c}$, single dot charging energy $U$, and the chemical potential local to the $j$th dot, $\mu_j\equiv\alpha_jV_j$, which is controlled by gate voltage $V_j$ with lever arm $\alpha_j$.  As in the main text, we define $\Delta = \mu_1-\mu_2$.  Then, for $\Delta\gg 0$, we have the approximate result
\be
J \approx
\sqrt{\ts{t}{c}^2+\frac{(U-|\Delta|)^2}{4}}-\frac{U-|\Delta|}{2},
\ee
which approaches $\ts{t}{c}^2/(U-|\Delta|)$ for $0\ll |\Delta|\ll U$, the ``detuning regime." In the SOP regime ($0\sim |\Delta|\ll U$),
\be
J \approx \frac{1}{1-\Delta^2/U^2}\left[\sqrt{2\ts{t}{c}^2+\frac{U^2}{4}}-\frac{U}{2}\right],
\label{SOPgap}
\ee
which approaches $2\ts{t}{c}^2U/(U^2-\Delta^2)$ for $\ts{t}{c}\ll U$.  Both the detuning and SOP regimes have $J\propto \ts{t}{c}^2$, but the key difference is that the detuning regime has $J\propto 1/(U-\Delta)$ while the SOP regime has $J\propto 1/(U^2-\Delta^2)$, reducing the sensitivity to $\Delta$ near $\Delta=0$.  These expressions assume that the triplet energy is independent of detuning, which holds only in the absence of excited states.  The effects of excited states may be included by subtracting a similar expression for triplets from \refeq{SOPgap} where the location and size of their avoided crossing is appropriately shifted.  The excited states have very little effect in the detuning regime since the triplet avoided crossing is far detuned from the singlet avoided crossing.  For symmetric operation ($\Delta=0$), however, a more accurate estimate of $J$ is given by
\begin{equation}
J = \frac{(t_{\rm c}^{\rm s}) ^2 }{U^{\rm s }} - \frac{(t_{\rm c}^{\rm t}) ^2 }{U^{\rm t} }= \frac{(t_{\rm c}^{\rm s}) ^2 }{U^{\rm s} \frac{U^{\rm t}}{U^{\rm t}-U^{\rm s}}}  +  \frac{(t_{\rm c}^{\rm s}) ^2-(t_{\rm c}^{\rm t}) ^2}{U^{\rm t}},
\end{equation}
where $t_{\rm c}^{\rm (s,t)}$ and $U^{\rm(s,t)}$ are the tunnel coupling and charging energy for the singlet and triplet states, respectively.  The final term in this expression can be neglected if the tunnel couplings are the same between the triplet and the singlet states.  This results in an effective rescaling of the charging energy $U$ by the percentage difference between the singlet and triplet charging energies.  (Note in the expressions above we have assumed symmetric dots; a difference in $U$ between the dots amounts only to an offset of $\Delta$, and the SOP is not degraded.)

In the SOP, $\Delta\sim 0$ and the tunnel coupling is the relevant control parameter.  Understanding the insensitivity in this regime requires a model for the voltage dependence of $\ts{t}{c}$.  We use a variant of the WKB approximation developed in \cite{RauModifiedWKB}.  Full three-dimensional simulations of our device indicate that this one-dimensional approximation is appropriate, at least for shallow barriers. The primary insight of this method is that using Weber (as opposed to Airy) functions to derive the WKB connection criteria better describes the eigenstates inside the shallow smooth barriers. This model gives
\begin{equation}
t_c \propto  \sqrt{\exp (2 \varphi) + 1} - \exp(\varphi),
\end{equation}
where $\varphi$ is the integral of $\sqrt{2{\ts{m}{e}}(\phi(x)-E)}$ between the classical turning points of an approximate 1D potential $\phi(x)$ at energy $E$.  If the barrier is large, $\exp \left(2 \varphi\right) \gg 1$, this expression reduces to the standard WKB double well tunnel coupling of $t_c \propto \exp (-\varphi)$; however, in the low barrier limit, $t_{\rm c}$ asymptotes to a constant as the two electron system merges into a single two-electron quantum dot.  In our fit model, we take $\varphi$ to be a linear function of voltage, which is justified by simulation to be appropriate for sufficiently shallow barriers.  From the inset of Fig.~5 of the main text, we see this sub-exponential behavior at large $J$. This behavior is responsible for insensitivity being an increasing function of exchange energy at high $J$.  At low $J$, the reduction of insensitivity visible in Fig.~5 is not well modeled by these approximations, but we find the observed reduction to be consistent with 3D Poisson-Schr\"odingier simulations in the presence of disorder due to surface charge.

\begin{figure}[h]
\includegraphics[width=\columnwidth]{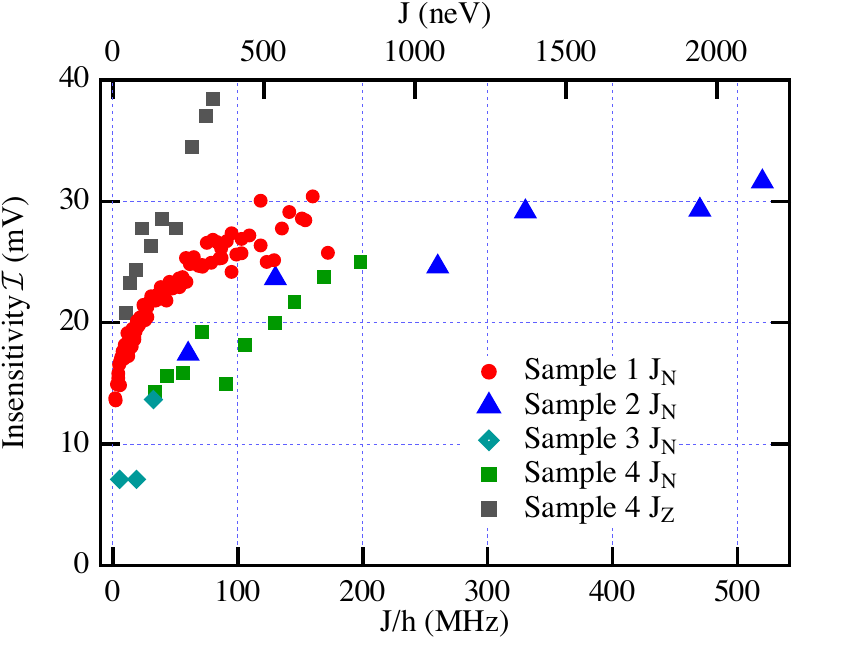}
\caption{Insensitivity as a function of exchange energy $J$ for four different samples.}
\label{IvsJ_multiple}
\end{figure}
The qualitative features of our model are observed in many different devices, as shown in Fig.~\ref{IvsJ_multiple}.  In this figure, Sample 1 is the device corresponding to Fig.~5 of the main text.  Small deviations in overall scale of insensitivity are explained by variations in the lever arms $\{\alpha_i\}$ that relate control voltages to the potential energies experienced by confined electrons.  These variations are due to both intentional and unintentional discrepancies in heterostructure and gate geometry between samples.  The plot also shows insensitivity for multiple exchange couplings in a single triple-dot device.  As described in Ref. \onlinecite{Eng2015etal},  rotations about the $\hat{z}$-axis or the $\hat{n}$-axis (rotated 120$^\circ$ from $\hat{z}$) are available in a triple dot depending on which pair of neighboring dots the tunnel barrier is lowered.   Figure~\ref{IvsJ_multiple} shows that the insensitivity of $J_Z$ and $J_N$ exchanges, measured in the same sample are qualitatively similar. \\

\section{Comparison to the Resonant Exchange Qubit}

Operating a triple-dot symmetrically to reduce sensitivity to charge noise has received recent attention in the context of the resonant exchange (RX) qubit  \cite{Taylor2013,Doherty2013,medfordprl}.  The RX qubit is an AC-controlled triple dot operated at a sweet spot along the axis of detuning between the outer dots: ${\Delta}_{\rm RX}\equiv\alpha(V_{\rm P3}-V_{\rm P1})$.   At $\ts\Delta{RX}=0$, a tunable gap due to simultaneous exchange across dots 1 and 2 and dots 2 and 3 provides continuous $\hat{z}$ rotations at frequency $\omega$.  Both the SOP and the RX-qubit sweet-spot have a gap which is second order in their respective tuning axes, as in \refeq{SOPgap}, owing to being tuned symmetrically with respect to two anticrossings.  Unlike SOP, however, the two anticrossings in the RX-qubit case correspond to tunneling into singlet states of different dots (1 and 3), and so the type of exchange coupling is different on either side of the $\ts\Delta{RX}=0$ sweet spot.  Resonant operation is accomplished by driving small oscillations in $\ts\Delta{RX}$ at frequency $\omega$,  resulting in Rabi oscillations at rate $\Omega$.

One advantage of the resonant exchange is that charge-noise-induced fluctuations in $\Omega$ become ineffective away from the resonant frequency $\omega$, reducing noise along this axis as $\omega$ is increased via tuning the tunnel barriers.  However, one is still subject to the full-bandwidth fluctuations in the gap itself.  In this sense, a comparable insensitivity measure for the RX qubit is the ratio of $\Omega$ to the voltage gradient of the gap.  As this gap has nearly equivalent properties to that in the SOP, the main difference is the available size of $\Omega$.  A key operating principle in RX is to keep $\Omega$ small in comparison to the gap $\omega$.  Otherwise control becomes complicated, either because the rotating wave approximation is violated or because the microwave control must compensate for the second-order variation of $\omega$ with $\ts\Delta{RX}$.  This forces a tradeoff between speed and accuracy for RX qubits that is not required of the SOP method; in fact,  as discussed above, the SOP performs better at faster Rabi rates.   Furthermore, reminiscent of DC-controlled triple-dot analysis in \cite{Fei2015}, one must leave the sweet spot of a different detuning parameter ${\Delta}'_{\rm RX}\equiv\alpha\left[\left(V_{\rm P1}+V_{P_3}\right)/2-V_{P_2}\right]$ to drive single qubit rotations, introducing first-order sensitivity of the gap to noise in ${\Delta}'_{\rm RX}$ during gate operations.

\section{Calibration of the Symmetric Operation Axis}

\begin{figure}[h]
\includegraphics[width=\columnwidth]{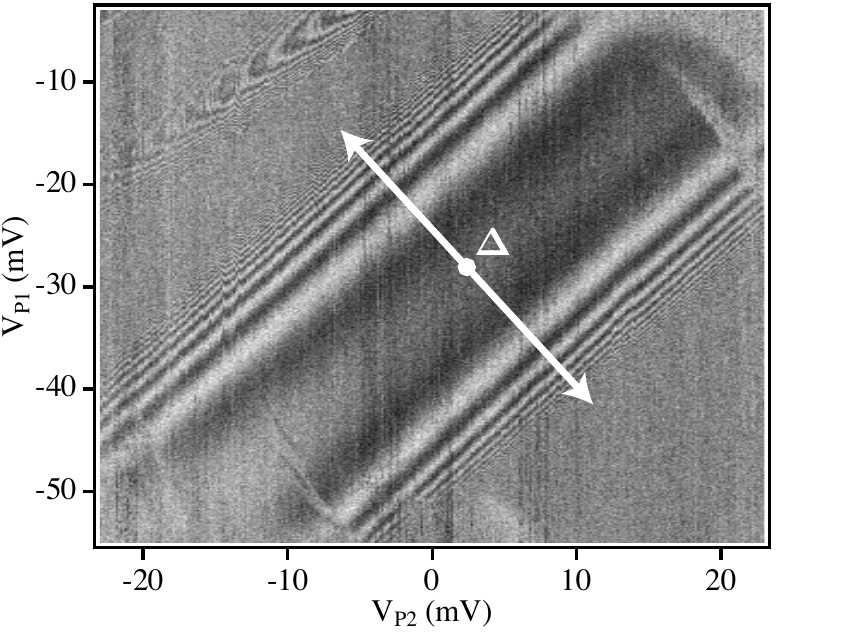}
\caption{Singlet probability as a function of $V_{\rm P1}$ and $V_{\rm P2}$ gate bias for a fixed $V_{\rm X1}$.   The (1,1) charge cell is observed with (0,2) and (2,0) charge boundaries on the lower right and upper left, respectively.  We observe coherent $J_n$ oscillations developing parallel to the (0,2) and (2,0) boundaries as the exchange energy rapidly increases (and the measured singlet population more rapidly oscillates) along the $\Delta$ detuning axis (drawn in white).}
\label{Calibration}
\end{figure}

As discussed in the main text, we apply voltages of the form ${\bf V}={\bf V}_0+\hat{{\bf u}}_{\rm detuning}({\Delta}/{\alpha}) + \hat{{\bf u}}_{\rm exchange}\tilde{V}_{\rm X1}$ where unit vectors $\hat{{\bf u}}_{\rm detuning}$ and $\hat{{\bf u}}_{\rm exchange}$ define the detuning and symmetric axes respectively, and $\vec{V}_0$ is a vector of the DC gate biases.   In order to exchange-couple electrons at a symmetric operating point, we achieve the desired exchange energy by tuning the  double-dot an amount $\tilde{V}_{\rm X1}$ along $\hat{{\bf u}}_{\rm exchange}$ with $\Delta=0$.   To calibrate $\hat{{\bf u}}_{\rm exchange}$, which depends on $V_{\rm X1}$, $V_{\rm P1}$, and $V_{\rm P2}$, we first find the center of the unit cell in the $V_{\rm P1}-V_{\rm P2}$ plane for a fixed, small $V_{\rm X1}$.   Sweeping $V_{\rm P1}$ and $V_{\rm P2}$, at each pixel a singlet is prepared (and later measured) in a third dot following the procedure outlined in  \cite{Eng2015etal} and in the main text.  A combination of voltages are pulsed in the  $V_{\rm P1}-V_{\rm P2}$ plane, followed by an exchange evolution.   As shown in Fig. \ref{Calibration}, the (1,1) charge cell is clearly observed by this procedure.  We denote the center of the (1,1) cell in the $V_{\rm P1}-V_{\rm P2}$ plane as $(V_{\rm P1}^{\rm c},V_{\rm P2}^{\rm c})$, defining a point  $\vec{P}(V_{\rm X1})\equiv(V_{\rm P1}^{\rm c},V_{\rm P2}^{\rm c},V_{\rm X1})$ in the 3D space. Due to cross capacitances, $(V_{\rm P1}^{\rm c},V_{\rm P2}^{\rm c})$  varies with $V_{\rm X1}$, so we repeat this procedure finding the center of the unit cell for a sufficiently different, fixed $V_{\rm X1}'$.  The symmetric voltage vector is defined as $ \hat{{\bf u}}_{\rm exchange}\equiv [\vec{P}(V_{\rm X1})-\vec{P}(V_{\rm X1}')]/|\vec{P}(V_{\rm X1})-\vec{P}(V_{\rm X1}')|$.  
The ``detuning'' axis specified by the vector $\hat{{\bf u}}_{\rm detuning}$ is orthogonal to the (2,0) to the (0,2) charge boundaries, corresponding  to increasing the chemical potential of one dot while simultaneously lowering that of the other.  A so-called ``fingerprint", as seen in Fig. 2 of the main text, is the result of repeating this prepare/evolve/measure procedure in the 2D plane defined by the symmetric and detuning axes, as obtained from this calibration procedure.

\begin{figure}[h]
\includegraphics[width=\columnwidth]{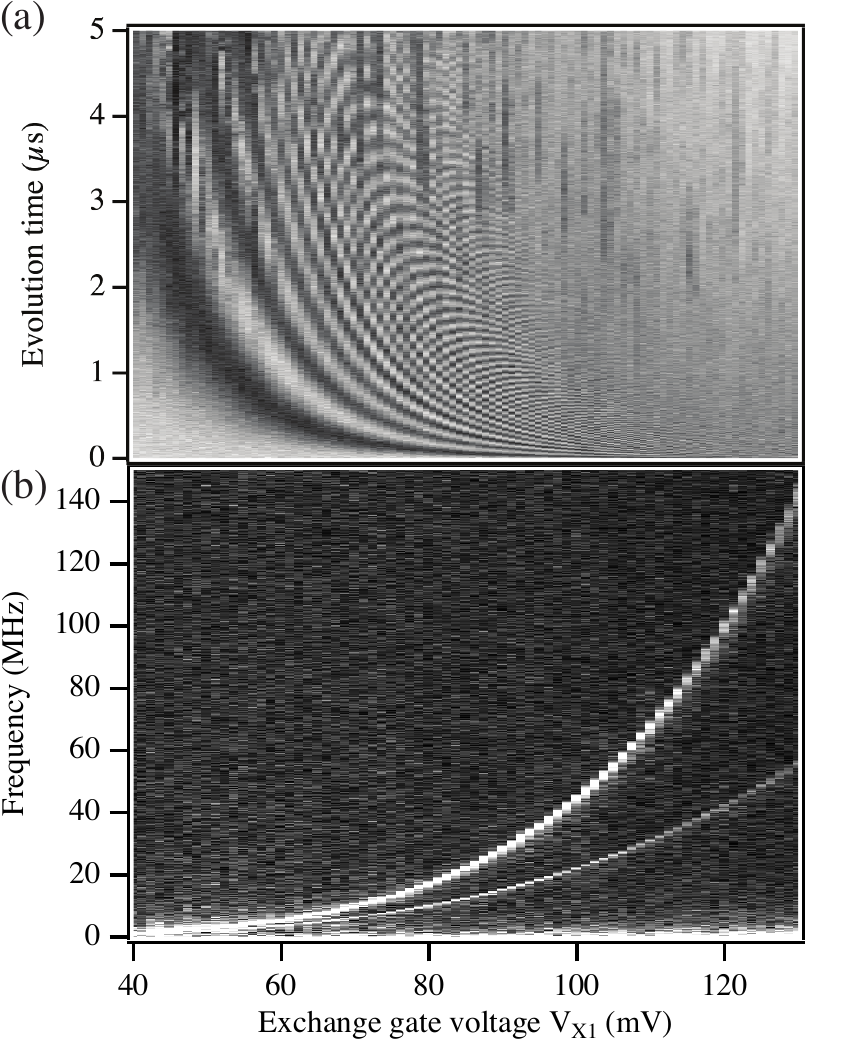}
\caption{(a) Rabi data as a function of exchange bias. (b) Fourier transform of (a), showing a clear signature of two energy curves, both of which are qualitatively similar functions of tunnel coupling.}
\label{TwoFreqs}
\end{figure}

\section{Elaboration of multiple fringes}

In some devices, Rabi and fingerprint data show faint signatures of a second fringe pattern with similar, but not perfectly overlapping, structure.    Extra fringes are faintly visible at large bias in  Fig. 2 in the main text and more evident in Rabi oscillation data in Fig. \ref{TwoFreqs}(a).
As shown in Fig. \ref{TwoFreqs}(b), these overlapping fringes are the result of incoherent superpositions of two distinct exchange energies that are clearly visible in the Fourier transform of the time-domain data.
Both sets of fringes are resolved by spin blockade in the same neighborhood of detuning space, suggesting that during operation, a double dot can be in an incoherent superposition of two distinct singlet/triplet pairs with charge transitions relatively close to each other, thus allowing simultaneous readout.
The two energy curves in \ref{TwoFreqs}(b) correspond to exchange energies for each of the two pairs, showing qualitatively similar (exponential) behavior with exchange gate bias.   From this data we obtain the exchange energy between each of the pairs but not the energy spacing from one pair to the other.    It is unclear whether the excited pair is a valley excited state, an orbital excited state, or some combination; this question will be the subject of future investigations. It is also unclear how the excited state is populated; however, thermal population of excited states may be significant for devices with relatively  small valley splitting.